\documentclass[10pt]{article}
\begin{document}

\begin{center}
{\large \bf Evaluating the Root-Mean-Squared of a Sinusoidal Signal Without Using Calculus}
\end{center}
\vspace{0.1in}

\begin{center}

{Rajesh R. Parwani\footnote{Email: parwani@nus.edu.sg}}

\vspace{0.2in}

{Department of Physics and\\}
{University Scholars Programme,\\}
{National University of Singapore,\\}
{Kent Ridge, Singapore.}

\vspace{0.2in}
%30 August 2004 \\ Revised 09 December 2004 \\ Revised 15 March 2005\\
%\maketitle
\end{center}
\vspace{0.1in}
\begin{abstract}
The frequently mentioned root-mean-squared value of an alternating voltage is derived without using  calculus.

\end{abstract}
%\pacs{}

\vspace{0.2in}

\section*{Alternating Voltages}

Students are taught that commercial electric power is available in two common forms, alternating and direct, but that it is alternating voltages that are supplied on large scales, such as to homes. When an alternating voltage given by  
\begin{equation}
V(t) = V_0  \sin(\omega t)
\end{equation} 
is applied across a resistor $R$, for example the coil of a water heater, heat is generated at a rate 
\begin{equation}
P(t) = { V^2(t) \over R} = { V_{0}^2 \over R}  \sin^2(\omega t) \, . \label{inst}
\end{equation} 
Since practical applications involve time scales much larger than $1/ \omega$, it is the average of the instantaneous power (\ref{inst}) over many cycles that is of interest. For the simple periodic function above, this is equivalent to asking for the average over one cycle. Textbooks quote this average as,
 \begin{equation}
<P> =  { V_{0}^2 \over  2 R} \, . \label{avepow}
\end{equation} 
This then motivates one to define the root-mean-squared voltage, given by $V_{rms} \equiv V_0 / \sqrt{2}$.
The root-mean-squared voltage is the constant voltage that would lead to the same heating effect as the alternating power averaged over many cycles.   

The ubiquitous $1/2$ in the mean-squared value (\ref{avepow}) can be obtained through an explicit integration. However it can also be derived simply by using two common trigonometric facts.
The first is that the functions $\sin (\theta)$ and $\cos(\theta)$ are periodic over the interval $0 \le \theta \le 2 \pi$ and they become identical when one of them is shifted  by $\pi/2$ relative to the other. Therefore the averages of $\sin^2 (\theta)$ and $\cos^2 (\theta)$ must be identical over one complete cycle. We also have the identity
\begin{equation}
\sin^2 (\theta) + \cos^2 (\theta) = 1 , \label{ident}
\end{equation} 
which, on taking the average over one cycle, gives 
\begin{equation}
2 < \sin^2 (\theta ) > = 1 , \label{result}
\end{equation} 
leading to desired result.

Students might wish for a more explicit, but non-calculus, discussion on what it means to take the average of an equation such as (\ref{ident}). Although the  interval $0 \le \theta \le 2\pi$ is continuous, one can evaluate the average of a function by sampling its value at a large number, $N$, of uniformly distributed points, and then taking the limit $N \to \infty$. Let $\theta_i$ denote a typical point, then the left-hand-side of (\ref{ident}) gives    
\begin{eqnarray}
< \sin^2 (\theta) + \cos^2 (\theta) > & \equiv & {1 \over N}  \sum_{i=1}^{i=N} ( \sin^2 (\theta_i) + \cos^2 (\theta_i) ) \\
&=&  {1 \over N}  \sum_{i=1}^{i=N} \sin^2 (\theta_i) +  {1 \over N}  \sum_{i=1}^{i=N}    \cos^2 (\theta_i)  \\
&=& {2 \over N}  \sum_{i=1}^{i=N} \sin^2 (\theta_i) \, ,
\end{eqnarray} 
which is the left-hand-side of (\ref{result}). The average of the right-hand-side of (\ref{ident}) similarly gives the right-hand-side of (\ref{result}).

An alternative non-calculus evaluation of the mean-squared value may be found in \cite{yf}. There authors adopt the trigonometric double-angle formula to write $\sin^2(\theta)$ in terms of  $\cos (2 \theta)$ and then use the fact that the average of a sinusoidal function is zero over one cycle. Arguably, the identity (\ref{ident}) is easier to recall than the double-angle formula. Indeed, in any right-angled triangle, which is used in elementary definitions of trigonometric functions, (\ref{ident}) is clearly equivalent to the famous Pythagoras theorem.

\end{document}